\documentclass[a4paper,11pt,twoside]{article}
\usepackage{vmargin}
\usepackage{natbib}
\usepackage{url}
\setmarginsrb{2cm}{2cm}{2cm}{2cm}{10mm}{10mm}{10mm}{10mm}
\usepackage[all]{nowidow}

\usepackage[breaklinks=true]{hyperref}
\usepackage{breakcites}

\usepackage{latexsym,amssymb,lastpage}
\usepackage{graphicx,amsfonts}
\usepackage{times,mathptmx,bm,amsmath}
\usepackage{dcolumn}

\usepackage{fancyhdr}
\usepackage{lastpage}
\fancypagestyle{title}{%
  \setlength{\headheight}{22pt}%
  \fancyhf{}% No header/footer
  \fancyfoot[R]{\scriptsize \copyright The authors 2014. Published by Oxford University Press on behalf of the Institute of Mathematics and its Applications. All rights reserved.}% Page number in Centre of footer
  \fancyhead[L]{\textsl{IMA Journal of Applied Mathematics} (2015) \textbf{80}, \thepage--\pageref{LastPage}\\
                doi:10.1093/imamat/hxu020\\ Advance Access publication on 6 June 2014}
  \fancyhead[R]{}
}%

\begin{document}
\title{\bfseries On the method of strained parameters for a KdV type of equation with exact dispersion property}
\author{{\sc Natanael Karjanto}\\[2pt]
Department of Mathematics, University College\\
Sungkyunkwan University, Natural~Science~Campus\\
2066 Seobu-ro, Jangan-gu, Suwon 16419, Gyeonggi-do\\
Republic of Korea\\ {\small \url{natanael@skku.edu}} \\[6pt]
%Department of Mathematics, School of Science and Technology, Nazarbayev University, 53~Kabanbay Batyr Avenue, Astana, 010000, Kazakhstan\\ {\rm natanael.karjanto@nu.edu.kz}\\[6pt]
%\rm [Received on 16 March 2012; accepted on 10 April 2014]\\[6pt]
}
%\pagestyle{headings}
%\markboth{N. KARJANTO}{\rm ON THE METHOD OF STRAINED PARAMETERS FOR A KDV TYPE OF EQUATION}
\date{\small }
\maketitle
\thispagestyle{title}

\begin{abstract}
\setlength{\parindent}{0pt}
This paper provides an alternative methodology for analysis of three-wave interactions under the exact dispersion relation associated with gravity waves in fluid of intermediate depth. A Korteweg-de Vries type of equation with exact dispersion property is adopted as the governing equation for unidirectional wave packet evolution. Following the idea from Zakharov's seminal paper (Zakharov, V. E. (1968) Stability of periodic waves of finite amplitude on the surface of a deep fluid. \textit{Journal of Applied Mechanics and Technical Physics}, {\bf 9}, 190--194), the equation is transformed from the spatial--temporal domain to the wavenumber--temporal domain. The solution of the transformed equation is sought using the perturbation theory, for which the ansatz is expressed in the form of a regular expansion in the increasing order of a small parameter. After implementing the na\"{i}ve perturbation method, due to nonlinear mode generation and particular combinations of wavenumbers, the third-order solution contains spurious secular growth terms which appear as a consequence of resonant interaction and nonlinear mode generation. These spurious secular growth terms can be prevented by implementing the method of strained parameters for which nonlinear dispersion relation terms are produced for particular combination of wavenumbers.\\

\end{abstract}

\section{Introduction}

In this paper, a Korteweg-de Vries (KdV) type of equation with exact dispersion property derived by \cite{vanGroesen98} and \cite{Cahyono02} is revisited. This type of equation has a slight difference in dispersion relation compared with the well-known classical KdV equation derived by \citet{KdV1895}. The dispersion relation of the classical KdV equation in the Boussinesq approximation is given by $\Omega(k) = k - \frac{1}{3!} k^3$, which is a third order approximation to the exact dispersion relation derived from fully nonlinear water waves equation given by $\Omega(k) = \sqrt{k\, \tanh\, k}$ \citep{Dean91,Debnath94}. The classical KdV equation is a good mathematical model for shallow-water wave propagation, i.e., for sufficiently long waves or small wavenumbers $k$. On the other hand, the KdV equation with exact dispersion relation is more appropriate mathematical model for surface gravity waves in fluid of intermediate depth. Furthermore, this latter equation has a special advantage, in particular if relatively short wavelengths are to be considered for modeling and application purposes in hydrodynamic laboratories \citep{vanGroesen98}. The model describes the unidirectional surface gravity waves evolution for which there is a balance between dispersion and nonlinearity. It is a special case of a novel equation derived using variational methods of Hamiltonian formulation for surface waves, named as the AB equation by the authors themselves \citep{vanGroesen07}. Recently, it has been shown that this AB equation provides an accurate description for unidirectional surface wave packet evolution \citep{vanGroesen10}. The authors illustrated the accuracy of the model in two different cases: for steady periodic Stokes' waves and for severely distorting running downstream bichromatic wave packets.

Although the theory of resonant interactions for surface gravity waves in a more general context is well established, this paper offers an alternative methodology for analysis of three wave interactions under the exact dispersion associated with gravity waves in fluid of intermediate depth. The existence of resonant interactions among surface gravity water waves was first observed by~\cite{Phillips60,Phillips61}. He showed that three-wave resonance with quadratic interactions is impossible while four-wave resonance with cubic nonlinearity is essential since the interactions provide a dominant coupling between modes. Several early works on the theory of resonant interactions amongst others are presented by~\cite{Benney62}, \cite{LH62} and~\cite{Hasselmann62,Hasselmann63}. The behaviour of wave group interaction in water of intermediate depth has been examined by~\cite{Bird01}. The interested readers may consult a comprehensive account of theory and experiment on wave interaction phenomena, both in fluids at rest and in shear flows, in the book by~\cite{Craik85}.

The focus of this paper is to find semi-analytical solution of the KdV equation with exact dispersion relation by implementing the perturbation theory~\citep{Nayfeh73,Kevorkian81,Hinch91}. Other alternative methods, such as the Adomian decomposition method, may also be implemented to obtain a series solution of the equation~\citep{Adomian94,Wazwaz09}. It has been shown that Adomian decomposition method is closely related to perturbation method~\citep{Sadat10}, where the author illustrated them in some heat conduction problems. More than a decade ago, finding semi-analytic solution of the KdV equation with exact dispersion relation by implementing the pertubation theory has been attempted by Cahyono in his PhD thesis~\citep{Cahyono02}. The author, however, did not attempt to make any transformation of the governing equation from the spatial--temporal ($xt$) domain to the wavenumber--temporal $(kt)$ domain. In this paper, we adopt the idea from~\cite{Zakharov68} by performing such transformation by applying the Fourier transform with respect to the wavenumber variable~$k$. As a consequence, the original governing equation of partial differential equation is reduced to an ordinary differential equation that depends only on the temporal variable~$t$. After implementing the na\"{i}ve perturbation method, also known as the Stokes expansion method, it is obtained that depending on the wavenumber combination, the third-order solution contains secular growth terms which appear to be the consequence of resonant interaction and nonlinear mode generation. Due to this perturbation series approach, the rise of a spurious secular variation of the solution is in contradiction with the periodic behaviour of the waves. By improving the method of perturbation, the occurrence of spurious secular growth terms can thus be prevented.

One of the improved perturbation techniques is called the method of strained parameters, which is also known as the Lindstedt--Poincar\'{e} method \citep{Nayfeh73,Dingemans97}. Another similar technique to remove spurious secular growth terms with a subtle difference to the method of strained parameters is known as the renomalization technique. This technique involves a variable substitution applied at the end result of the na\"{i}ve perturbation method. The method of strained parameters, on the other hand, utilizes a variable transformation at an earlier stage to tackle resonant driving terms as a cause for spurious secular growth. Using this method, both the transformed wave field and the dispersion relation are expressed and expanded in the powers of a small, positive parameter $\epsilon$, most often represents the wave steepness. Furthermore, by requiring the vanishing of spurious secular growth terms, it leads to expressions for the $n^{\textmd {th}}$-order, nonlinear dispersion relations $\Omega_n$, $n \in \mathbb{N}$. This method was initially developed by Lindstedt and later Poincar\'{e} showed that the expansions obtained earlier by Lindstedt's technique are asymptotic, thus, coined the name of the technique as the Lindstedt--Poincar\'{e} method. Although the method has been discovered for more than a century, it is still as effective as it was back then. In this paper, we combine both the methods of na\"{i}ve perturbation and strained parameters to derive a semi-analytic solution of the KdV equation with exact dispersion relation.

As mentioned earlier, the transformation of the governing equation from the $xt$ domain to the $kt$ domain is implemented by adopting the approach from Zakharov's seminal paper~\citep{Zakharov68}. Nonetheless, there is an essential difference in term of employing the governing equations. The Zakharov integral equation, also known as Zakharov's equation, is derived from the fully nonlinear equation for water waves. The governing equations for wave motion are given by the Laplace equation for wave potential, kinematic and dynamic boundary conditions at the free surface and a kinematic boundary condition at the bottom. The Zakharov integral equation describes the slow temporal evolution of the dominant Fourier components of a weakly nonlinear surface gravity wave field on deep water~\citep{Dingemans01}. On the other hand, the KdV equation with exact dispersion relation is a unidirectional model for surface wave packet propagation for intermediate water depth. This equation is derived from Boussinesq equations by implementing a unidirectionalization procedure. Boussinesq equations are derived from the fully nonlinear water wave equations governed by the Laplace equations for the fluid potential and dynamic and kinematic boundary conditions at the free surface~\citep{Debnath94,Cahyono02}.

Using the assumption of narrow-band spectra, expressing the surface wave packet as a superposition of the first- and the second-order harmonic intermediate waves and non-harmonic long wave and implementing the method of multiple time-scale, the corresponding complex-valued wave packet amplitude satisfies temporal and spatial nonlinear Schr\"{o}dinger equations, depending on the choice of dynamic frame of reference variables~\citep{vanGroesen98,Karjanto06}. The latter equation has been chosen as a mathematical model for freak wave generation in a hydrodynamic laboratory and several families of solutions exhibit large amplitude increase, phase singularity, wavefront dislocation and frequency downshift, as shown both theoretically and experimentally~\citep{Karjanto02,Huijsmans05,vanGroesen05,vanGroesen06,Andonowati07,Karjanto07,Karjanto08,Karjanto09,Karjanto10}. Interested readers are also encouraged to consult~\cite{Dysthe99,Henderson99,Osborne00,Osborne01,Onorato01,Onorato06,Pelinovsky00,Kharif03,Kharif09,Zakharov06,Solli07,Dyachenko08,Shrira10} for more insightful discussions on freak waves.

This paper is organized as follows. In the following section, we will introduce the KdV type of equation with exact dispersion property, taking its Fourier transform and implementing the na\"{i}ve perturbation method to obtained a transformed equation in the Fourier domain. Section~\ref{strained} explains the method of strained parameters, i.e. an effective procedure to get rid spurious secular growth terms occurred due to resonant interaction and nonlinear mode generation employed by the na\"{i}ve perturbation method. Section~\ref{valid} presents a validity analysis of the second-order nonlinear dispersion relation and provides potential wavenumber thresholds for the transition from the shallow-water wave to intermediate-water wave and from the intermediate-water wave to deep-water wave. Finally, some conclusions are presented at the end of the paper.

\section{Na\"{i}ve perturbation method} \label{naive}

Consider a nonlinear dispersive wave equation for a mathematical model of wave packet propagation. In this study, we adopt the KdV type of equation with exact dispersion property and an imposed initial condition\footnote{An initial value problem is considered in this paper. A boundary value problem or a signalling problem can also be studied by imposing an initial signal at a boundary instead of an initial condition.} \citep{vanGroesen98}. It reads
\begin{equation}
\partial_{t} \eta  + i\,\Omega(-i\,\partial_{x})\, \eta + \frac{3}{4} \partial_{x}\,(\eta^2) = 0, \qquad \eta(x,0) = u_1(x), \qquad x \in \mathbb{R}, \qquad t \geq 0. \label{KdV}
%\epsilon u^{(1)}(x)
\end{equation}
The derivation of this equation using variational structure as well as its relationship with the classical KdV equation are explained by~\cite{vanGroesen94}.
In this equation, $\eta(x,t)$ denotes surface wave elevation, $\Omega$ is a skew-symmetric preudo-differential (Fourier-integral) operator related to the exact linear dispersion relation, in non-normalized form given as $\Omega(k) = \sqrt{g k \, \tanh (k h)}$, where $g$ is the acceleration of gravity and $h$ is the water depth~\citep{Lamb94}.
According to Airy (linear) wave theory, for shallow-water wave, where $kh \ll 1$, the linear dispersion relation reduces to $\Omega(k) = k \sqrt{gh}$ and for deep-water wave, where $kh \gg 1$, the linear dispersion relation reduces to $\Omega(k) = \sqrt{gk}$~\citep{Phillips77,Dingemans97}. In normalized form, the exact linear dispersion relation reads $\Omega(k) = \sqrt{k \, \tanh\, k}$.
For dispersive wave equation, $\Omega(k)$ is a real-valued function for all $k \in \mathbb{R}$ and is an odd function, i.e. $\Omega(-k) = -[\Omega(k)]^\ast = - \Omega(k)$.
The initial condition $u_1(x) = {\cal O}(\epsilon)$, where $0 < \epsilon \ll 1$ and $u_1 \in L^{1}(\mathbb{R})$.

Implementing Zakharov's approach~\citep{Zakharov68} by applying the Fourier transform with respect to the spatial domain, the governing equation~\eqref{KdV} is transformed from the spatial--temporal ($xt$) domain into the wavenumber--temporal ($kt$) domain. As a consequence, the initial value problem is reduced from a partial differential equation (PDE) to an ordinary differential equation (ODE) with respect to time variable~$t$. We also adopt the following convention and notation to define the Fourier transform, its inverse and the convolution of two functions, respectively:
\begin{eqnarray*}
  \hat{\eta}(k,t) &=& {\cal F}\{ \eta(x,t) \} = \int_{-\infty}^{\infty} \eta(x,t) e^{ikx} \, dx \\
  \eta(x,t) &=& {\cal F}^{-1} \{\hat{\eta}(k,t) \} = \frac{1}{2\pi}\int_{-\infty}^{\infty} \hat{\eta}(k,t) e^{-ikx} \, dk \\
  (\hat{\eta}_1 \ast \hat{\eta}_2)(k,t) &=& \int_{-\infty}^{\infty} \hat{\eta}_1 (k - \kappa,t) \, \hat{\eta}_2 (\kappa,t) \, d\kappa = \int_{-\infty}^{\infty} \hat{\eta}_1 (\kappa,t) \, \hat{\eta}_2 (k - \kappa,t) \, d\kappa.
\end{eqnarray*}

Applying the Fourier transform to \eqref{KdV} yields
\begin{equation}
  \partial_{t}\hat{\eta} + i\,\Omega(k)\hat{\eta} + \frac{3ik}{8 \pi} \,(\hat{\eta} \ast \hat{\eta}) = 0, \qquad \hat{\eta}(k,0) = \hat{u}_1(k), \qquad t \geq 0. \label{FTKdV}
%  \epsilon \hat{u}^{(1)}(k)
\end{equation}
Introduce the transformation solution $\hat{u}(k,t) = \hat{\eta}(k,t)\,e^{\,i\Omega(k)t}$, where $\hat{\eta}(k,t)$ is a solution of \eqref{FTKdV}. It turns out that this transformation is rather useful since the solution to the series expansion would have shorter expressions. The transformed KdV equation with exact dispersion relation \eqref{FTKdV} now turns to
\begin{equation}
\partial_{t}\hat{u} + \frac{3ik}{8\pi}\,e^{\,i\Omega(k)t} \left(\hat{u}\,e^{-\,i\Omega(k)t} \ast
\hat{u}\,e^{\,-i\Omega(k)t} \right) = 0. \label{TFTKdV}
\end{equation}
Write the transformation solution $\hat{u}(k,t)$ as a series expansion in $\epsilon$:
\begin{equation*}
  \hat{u}(k,t) = \sum_{n = 1}^\infty \hat{u}_n (k,t) = \sum_{n = 1}^\infty \epsilon^n \hat{u}^{(n)}(k,t)
\end{equation*}
and implement the regular (na\"{i}ve) perturbation method to solve \eqref{TFTKdV}~\citep{Nayfeh73,Kevorkian81}.
Collecting the terms of like orders in $\epsilon$, setting them equal to zero, a set of differential equations that depends only on the time variable $t$ is obtained.
Once the solution for the lowest order equation $\hat{u}^{(1)}$ is found, higher order solutions $\hat{u}^{(2)}, \hat{u}^{(3)}, \dots$ are solved by integrating these differential equations with respect to $t$ and the solutions are expressed in terms of integrals which appear due to convolution terms of the ODEs.
The list of equations up to the third order is given as follows:
\begin{eqnarray}
  \textmd{$\cal{O}$}(\epsilon) : &\quad& \partial_{t}\hat{u}^{(1)} = 0 \\
  \textmd{$\cal{O}$}(\epsilon^2) : &\quad& \partial_{t}\hat{u}^{(2)}  + \frac{3ik}{8 \pi} e^{i \Omega(k) t}
  \left( \hat{u}^{(1)} e^{-i\Omega(k) t} \ast \hat{u}^{(1)} e^{-i\Omega(k) t}\right) = 0 \label{secondoe}\\
  \textmd{$\cal{O}$}(\epsilon^3) : &\qquad& \partial_{t}\hat{u}^{(3)} + \frac{3ik}{4\pi} e^{i\Omega(k) t}
  \left( \hat{u}^{(1)} e^{-i\Omega(k) t} \ast \hat{u}^{(2)} e^{-i\Omega(k) t} \right) = 0. \label{thirdoe}
\end{eqnarray}
The lowest order equation solves $\hat{u}^{(1)}(k,t) = \hat{u}^{(1)}(k)$.
The second order solution $\hat{u}^{(2)}$ is acquired by integrating the second order equation~\eqref{secondoe} with respect to $t$ by taking the lower boundary of integration as $t = 0$. For $j \in \mathbb{N}$, it reads
\begin{equation}
\hat{u}^{(2)}(k,t) = \left\{ \begin{array}{ll}
         {\displaystyle -\frac{3k}{8\pi} \int_{-\infty}^\infty \frac{\hat{v}_{0j}}{\psi_{0j}} \left(e^{i\psi_{0j}t} - 1 \right) \, dk_j}, & \quad \mbox{for $k_j \neq k$}\\
        0, & \quad \mbox{for $k_j = k$} \end{array} \right. \label{2ndsolu}
\end{equation}
where
\begin{eqnarray*}
\hat{v}_{0j} &=& \hat{u}^{(1)}(k_j) \, \hat{u}^{(1)}(k - k_j) \\
\psi_{0j}      &=& \Omega(k) - \Omega(k_j) - \Omega(k - k_j).
\end{eqnarray*}
It is obvious that the restriction $k_j \neq k$, for $j \in \mathbb{N}$ is very essential to guarantee the existence of a non-trivial series solution.

To obtain the third order solution $\hat{u}^{(3)}$, simply integrate the third order equation~\eqref{thirdoe} with respect to the temporal variable $t$ and by employing $t = 0$ as the lower boundary of integration, it is found that for $k_1 \neq k$, $k_2 \neq k$, $k_1 + k_2 \neq 0$ and $k_2 - k_1 \neq k$, it reads
\begin{eqnarray}
\hat{u}^{(3)}(k, t) &=& \frac{9k}{64\pi^2} \int_{-\infty}^\infty \int_{-\infty}^\infty k_2 \frac{\hat{v}_{11}}{\psi_{11}} \left(\frac{e^{i \psi_{21}t }}{\psi_{21}} -
\frac{e^{i\psi_{02}t}}{\psi_{02}} +  C_{11} \right)\, d k_1 \, d k_2 \nonumber \\
&& + \, \frac{9k}{64\pi^2} \int_{-\infty}^\infty \int_{-\infty}^\infty (k - k_2) \frac{\hat{v}_{12}}{\psi_{12}} \left(\frac{e^{i \psi_{22}t }}{\psi_{22}} -
\frac{e^{i\psi_{02}t}}{\psi_{02}} + C_{12} \right) \, d k_1 \, d k_2 \label{3rdsolu}
\end{eqnarray}
where
\begin{eqnarray*}
C_{11} &=& \frac{1}{\psi_{02}} - \frac{1}{\psi_{21}}  \\
C_{12} &=& \frac{1}{\psi_{02}} - \frac{1}{\psi_{22}}
\end{eqnarray*}
\begin{eqnarray*}
\hat{v}_{11} &=& \hat{u}^{(1)}(k_{1}) \hat{u}^{(1)}(k_{2}) \hat{u}^{(1)}(k_{2} - k_{1}) \\
\hat{v}_{12} &=& \hat{u}^{(1)}(k_{1}) \hat{u}^{(1)}(k_{2}) \hat{u}^{(1)}(k - k_{1} - k_{2}) \\
\psi_{02} &=& \Omega(k) - \Omega(k_2) - \Omega(k - k_2) \\
\psi_{11} &=& \Omega(k_{2}) - \Omega(k_{1}) - \Omega( k_{2} - k_{1}) \\
\psi_{12} &=& \Omega(k - k_{2}) - \Omega(k_{1}) - \Omega(k - k_{1} - k_{2}) \\
\psi_{21} = \psi_{02} + \psi_{11} &=& \Omega(k) - \Omega(k_{1}) - \Omega(k_{2} - k_1) - \Omega(k - k_{2}) \\
\psi_{22} = \psi_{02} + \psi_{12} &=& \Omega(k) - \Omega(k_{1}) - \Omega(k_{2}) -  \Omega(k - k_{1} - k_2).
\end{eqnarray*}
For either $k_1 + k_2 = 0$ or $k_2 - k_1 = k$, the third order solution produces a spurious secular growth term that grows linearly in time, given as follows:
\begin{equation}
  \hat{u}^{(3)}(k, t) = \left\{
                          \begin{array}{ll}
                          {\displaystyle \frac{9k\, t}{64\pi^2} \int_{-\infty}^\infty \int_{-\infty}^\infty k_2 \frac{\hat{v}_{11}}{\psi_{11}} \, d k_1 \, d k_2 + \textmd{non-secular terms}}, & \hbox{for} \; k_1 + k_2 = 0 \\
                          {\displaystyle \frac{9k\, t}{64\pi^2} \int_{-\infty}^\infty \int_{-\infty}^\infty (k - k_2) \frac{\hat{v}_{12}}{\psi_{12}} \,  d k_1 \, d k_2 + \textmd{non-secular terms}}, & \hbox{for} \; k_2 - k_1 = k.
                          \end{array}
                        \right.
\end{equation}
This na\"{i}ve perturbation method will break down when $t \sim 1/\epsilon^3$ since $\hat{u}_3(k, t) = \epsilon^3 \hat{u}^{(3)}$ will be of the same order of $\hat{u}_1$ and violate the uniformity of the convergence of the asymptotic expansion. This $t$~dependence in $u_3$ is known as spurious secular growth and arises whenever there is a resonance between $\hat{u}_2$ and $\hat{u}_3$. This resonance is a special case of nonlinear mode generation through nonlinearity and occurs when the wave modes of Fourier components are mixing up and generate other wave modes due to nonlinear effect.
This spurious secular growth is undesirable and can be prevented by improving the method of perturbation expansion. It is important to note that an identical spurious secular growth term will also appear even if one implements different techniques to acquire semi-analytical solution of the problem, for instance using the Adomian decomposition method~\citep{Adomian94,Wazwaz09}. In the following section, we will discus the method of strained parameters to show that a bounded solution can be obtained indeed.

\section{The method of strained parameters} \label{strained}

The appearance of resonant terms can be prevented by implementing the method of strained parameters. This method suggests to write the dispersion relation $\Omega(k)$ as $\Omega(k; \epsilon)$ and expand it in the powers of $\epsilon$, expressed as follows:
\begin{equation}
  \Omega(k; \epsilon) = \sum_{n = 0}^\infty \epsilon^n \Omega_{n}(k) = \Omega_{0}(k) + \epsilon\,\Omega_{1}(k) + \epsilon^2\,\Omega_{2}(k) + \cdots, \label{ndr0}
\end{equation}
where the lowest order term $\Omega_0(k) = \sqrt{k \tanh k} = k \sqrt{\tanh k/k}$ is the exact, linear dispersion relation in normalized form given in Section~\ref{naive} and the higher order terms represent the $n^{\textmd{th}}$-order nonlinear dispersion relations. Substituting the series expansion \eqref{ndr0} into the transformed KdV equation with exact dispersion relation~\eqref{FTKdV}, we obtain a set of ODEs according to the order of $\epsilon$. The lowest order equation now reads
\begin{equation}
  \textmd{$\cal{O}$}(\epsilon) : \quad \partial_{t}\hat{\eta}^{(1)} + i\,\Omega_{0}(k)\hat{\eta}^{(1)} = 0.
\end{equation}
The solution of this equation is $\hat{\eta}^{(1)}(k,t) = \hat{\eta}^{(1)}(k,0) e^{\,-i\,\Omega_{0}(k)t}$. Using a similar transformation introduced in the previous section,  $\hat{u}^{(j)}(k,t) = \hat{\eta}^{(j)}(k,t) e^{\,i\Omega_{0}(k)t}$, $j \in \mathbb{N}$,
a time independent solution for the lowest order equation is readily obtained: $\hat{u}^{(1)}(k,t) = \hat{\eta}^{(1)}(k,0) = \hat{u}^{(1)}(k)$.

Collecting the second-order terms gives us an ODE with the nonlinearity containing the self convolution of the first-order solution $\hat{\eta}^{(1)}$ and an additional term containing the product of the first-order nonlinear dispersion relation $\Omega_1(k)$ and the first-order solution $\hat{\eta}^{(1)}$, given as follows:
\begin{equation}
\textmd{$\cal{O}$}(\epsilon^2) : \quad \partial_{t}\hat{\eta}^{(2)} + i\,\Omega_{0}(k)\hat{\eta}^{(2)} + i\,\Omega_{1}(k)\hat{\eta}^{(1)} +
\frac{3ik}{8 \pi} \left( \hat{\eta}^{(1)} \ast \hat{\eta}^{(1)} \right) = 0.
\end{equation}
Applying again the transformation introduced earlier and integrating with respect to the temporal variable $t$, we obtain the second-order solution $\hat{u}^{(2)}(k,t)$:
\begin{equation}
\hat{u}^{(2)}(k,t) = \left\{ \begin{array}{ll}
         {\displaystyle - i \Omega_1(k) u^{(1)}(k) t - \frac{3k}{8\pi} \int_{-\infty}^\infty \frac{\hat{v}_{0j}}{\Psi_{0j}} \left(e^{i\psi_{0j}t} - 1 \right) \, dk_j}, & \quad \mbox{for $k_j \neq k$}\\
        - i \Omega_1(k) u^{(1)}(k) t , & \quad \mbox{for $k_j = k$} \end{array} \right.
\end{equation}
where
\begin{equation*}
\Psi_{0j}      = \Omega_0(k) - \Omega_0(k_j) - \Omega_0(k - k_j).
\end{equation*}
Since the first term of the solution produces an undesirable spurious secular growth in $t$ which will be the same order of $\hat{u}_1$ for $t \sim 1/\epsilon^2$, the first-order nonlinear dispersion relation is chosen to be zero, i.e. $\Omega_1(k) = 0$.

Collecting the third-order terms yields an ODE in $\hat{\eta}^{(3)}$ with nonlinear and nonhomogeneous terms that depend on the previous lower order solutions $\hat{\eta}^{(1)}$ and $\hat{\eta}^{(2)}$. It reads
\begin{equation}
\textmd{$\cal{O}$}(\epsilon^3) : \partial_{t}\hat{\eta}^{(3)} + i\,\Omega_{0}(k) \hat{\eta}^{(3)} + i\,\Omega_{2}(k)\hat{\eta}^{(1)} + \frac{3ik}{4\pi} \left( \hat{\eta}^{(1)} \ast \hat{\eta}^{(2)}\right) = 0.
\end{equation}
Applying again the transformation introduced earlier, we can write an expression for the ODE as follows:
\begin{eqnarray}
\partial_{t}\hat{u}^{(3)} &=& - i \Omega_2(k) \, \hat{u}^{(1)}(k)
+ \frac{9ik}{64\pi^2} \int_{-\infty}^{\infty}\int_{-\infty}^{\infty} k_2 \frac{\hat{v}_{11}}{\Psi_{11}} (e^{i \Psi_{21} t} - e^{i \Psi_{02} t}) \, dk_1 \, dk_2  \nonumber\\
&& + \; \frac{9ik}{64\pi^2} \int_{-\infty}^{\infty}\int_{-\infty}^{\infty} (k - k_2) \frac{\hat{v}_{12}}{\Psi_{12}} \, \left(e^{i \Psi_{22} t} - e^{i \Psi_{02} t} \right) \, dk_{1} \,dk_{2}
\end{eqnarray}
where
\begin{eqnarray*}
\Psi_{02} &=& \Omega_0(k) - \Omega_0(k_2) - \Omega_0(k - k_2) \\
\Psi_{11} &=& \Omega_0(k_{2}) - \Omega_0(k_{1}) - \Omega_0( k_{2} - k_{1}) \\
\Psi_{12} &=& \Omega_0(k - k_{2}) - \Omega_0(k_{1}) - \Omega_0(k - k_{1} - k_{2}) \\
\Psi_{21} = \Psi_{02} + \Psi_{11} &=& \Omega_0(k) - \Omega_0(k_{1}) - \Omega_0(k_{2} - k_1) - \Omega_0(k - k_{2}) \\
\Psi_{22} = \Psi_{02} + \Psi_{12} &=& \Omega_0(k) - \Omega_0(k_{1}) - \Omega_0(k_{2}) -  \Omega_0(k - k_{1} - k_2).
\end{eqnarray*}

In order to prevent the occurrence of spurious secular growth terms, the second-order nonlinear dispersion relation $\Omega_2(k)$ is chosen to satisfy the following conditions, depending on the combination of the wavenumbers $k_1$ and $k_2$:
\begin{itemize}
\item for $k_1 + k_2 = 0$
\begin{equation}
  \Omega_2(k) = \frac{9}{64 \pi^2} \frac{k}{\hat{u}^{(1)}(k) } \int_{-\infty}^{\infty} \frac{k_1\hat{U}_1(k_1)}{\Omega_0(2k_1) - 2 \Omega_0(k_1)} \, dk_1 \label{ndr1}
\end{equation}
where
\begin{equation*}
\hat{U}_1 (k_1) = \hat{u}^{(1)}(k_1) \hat{u}^{(1)}(-k_1) \hat{u}^{(1)}(-2 k_1).
\end{equation*}

\item for $k_2 - k_1 = k$
\begin{equation}
  \Omega_2(k) = \frac{9}{64 \pi^2} \frac{k}{\hat{u}^{(1)}(k) } \int_{-\infty}^{\infty} \frac{k_1\hat{U}_2(k_1)}{\Omega_0(2k_1) - 2 \Omega_0(k_1)} \, dk_1 \label{ndr2}
\end{equation}
where
\begin{equation*}
\hat{U}_2 (k_1) = - \hat{u}^{(1)}(k_1) \hat{u}^{(1)}(k + k_1) \hat{u}^{(1)}(-2 k_1).
\end{equation*}
\end{itemize}
To ensure the correspondence with the initial wave profile or wave signal which is a real-valued function, the dispersion relation is also regarded as an odd and a real valued function for all $k \in \mathbb{R}$, i.e. $\Omega(k) = -[\Omega(-k)]^\ast = -\Omega(-k)$. Note that this is also consistent with the linear dispersion relation for the classical KdV equation which is also an odd function.

The solution of the transformed KdV equation with exact dispersion relation \eqref{FTKdV} up to the third order term now reads
\begin{equation*}
  \hat{\eta}(k,t) = \left[\epsilon \hat{u}^{(1)}(k) + \epsilon^2 \hat{u}^{(2)}(k,t) + \epsilon^3 \hat{u}^{(3)}(k,t) \right] e^{-i\left[\Omega_0(k) + \epsilon^2 \Omega_2(k) \right]t}
\end{equation*}
where $\hat{u}^{(1)}$ is the Fourier transform of the initial condition $u_1(x)/\epsilon$, $\hat{u}^{(2)}$ and $\hat{u}^{(3)}$ are the Fourier transforms of the second and the third order solutions, \eqref{2ndsolu} and \eqref{3rdsolu}, respectively; $\Omega_0(k)$ is the exact linear dispersion relation and $\Omega_2(k)$ is the second order nonlinear dispersion relation \eqref{ndr1} or \eqref{ndr2}.
By taking the inverse Fourier transform, the solution of the KdV equation with exact dispersion type is readily obtained, up to the third order term is given as follows:
\begin{equation*}
  \eta(x,t) = \frac{1}{2\pi} \int_{-\infty}^{\infty} \left[\epsilon \hat{u}^{(1)}(k) + \epsilon^2 \hat{u}^{(2)}(k,t) + \epsilon^3 \hat{u}^{(3)}(k,t) \right] e^{-i\left(kx + \left[\Omega_0(k) + \epsilon^2 \Omega_2(k) \right]t \right)} dk + {\cal O}(\epsilon^4).
\end{equation*}
The following section discusses the validity analysis of the of the solution obtained by the method of strained parameters and a  potential wavenumber threshold for the transition from intermediate-water wave to deep-water wave.

\section{Validity analysis} \label{valid}
Theoretically, for any given initial condition $\eta(x,0)$, one can calculate its Fourier transform $\hat{u}^{(1)}(k)$ and the second-order nonlinear dispersion relation $\Omega_2$. In practice, however, there are constraints  need to be considered in order to find $\Omega_2$.
The exact linear dispersion relation $\Omega_0$ that appears on the denominator of the integrand of $\Omega_2$ should be understood as having different expressions as the integrand term is integrated with respect to the wavenumber $k_1$.

For shallow-water wave case (long wavelength) when $|k_1| \ll 1$, i.e. $k_1 = {\cal O}(\epsilon)$, $\Omega_0$ reduces to the linear dispersion for the classical KdV equation, thus the denominator of the integrand in~\eqref{ndr1} and~\eqref{ndr2} reduces to:
\begin{equation*}
\Omega_0(2k_1) - 2\Omega_0(k_1) \approx -k_1^3, \qquad \textmd{for} \quad |k_1| \ll 1.
%\left( 2k_1 - \frac{(2k_1)^3}{3!} \right) - \left(2k_1 - \frac{2}{3!} k_1^3 \right)  = -k_1^3, \qquad \textmd{for} \quad |k_1| << 1.
\end{equation*}
Thus, for shallow-water wave case, it is sufficient to take any initial condition such that with $\hat{u}^{(1)} = {\cal O}(1)$ to guarantee that $\Omega_2$ is integrable or the integrand of $\Omega_2 \in L^2(\mathbb{R})$. This is consistent with the imposed initial condition $\hat{u}_1 = \epsilon \hat{u}^{(1)}$.

On the other hand, for deep-water wave case (short wavelength) when $|k_1| \gg 1$, i.e. $k_1 = {\cal O}(\epsilon^{-1})$, the exact dispersion relation $\Omega_0$ reduces to the dispersion relation for deep-water wave case. Since $\tanh k_1 \approx 1$ for large $|k_1|$, the denominator of the integrand in~\eqref{ndr1} and~\eqref{ndr2} reduces to
\begin{equation*}
\Omega_0(2k_1) - 2\Omega_0(k_1) \approx \left( \sqrt{2} - 2 \right) \sqrt{k_1}, \qquad \textmd{for} \quad |k_1| \gg 1.
\end{equation*}
Thus, for deep-water wave case, it is necessary to impose the initial condition such that $\hat{u}^{(1)} = {\cal O}({k}^{-5/6})$ to guarantee that $\Omega_2$ is integrable or its integrand is in $L^2(\mathbb{R})$. This is almost consistent with the imposed initial condition of $\hat{u}_1 = {\cal O}(\epsilon)$ since now $\hat{u}^{(1)} = {\cal O}(\epsilon^{5/6}) \approx {\cal O}(1)$.

For intermediate-water wave case (finite depth), rationalizing the denominator of the integrand of $\Omega_2$, the integrals of~\eqref{ndr1} and~\eqref{ndr2} can be expressed as (in terms of order, after dropping the factor $-\frac{1}{4}$):
\begin{equation*}
\int \frac{\hat{U}_{1,2}(k_1)}{\tanh k_1} \left(1 + \frac{1}{\tanh^2 k_1} \right) \left[\Omega_0(2k_1) + 2 \Omega_0(k_1) \right] \, dk_1.
\end{equation*}
Defining the integrals in the expressions for $\Omega_2$ in~\eqref{ndr1} and~\eqref{ndr2} as $I_1$ and $I_2$, respectively, we have
\begin{eqnarray*}
I_{1,2} &=& \int_{-\infty}^{-k_d} \hat{U}_{1,2} \, {\cal O}(\sqrt{k_1}) dk_1 + \int_{k_d}^{\infty} \hat{U}_{1,2} \,  {\cal O}(\sqrt{k_1}) dk_1  \\
&& + \; \int_{-k_d}^{-k_s} \frac{k_1 \hat{U}_{1,2}}{\Omega_0(2k_1) - 2 \Omega_0(k_1)}  dk_1 + \int_{k_s}^{k_d} \frac{k_1 \hat{U}_{1,2}}{\Omega_0(2k_1) - 2 \Omega_0(k_1)} dk_1 \\
&& + \; \int_{-k_s}^{k_s} \hat{U}_{1,2} \,  {\cal O}(k_1^{-2}) dk_1
\end{eqnarray*}
where $k_s$ and $k_d$ are the wavenumber thresholds for the shallow-water wave and deep-water wave cases, respectively.
Furthermore, we have
\begin{equation*}
\left| \int_{-k_s}^{k_s} \hat{U}_{1,2} \,  {\cal O}(k_1^{-2}) dk_1 \right| \leq \int_{-k_s}^{k_s} \left| \hat{U}_{1,2} \,  {\cal O}(k_1^{-2}) \right| dk_1 \leq \int_{-\infty}^{\infty} \left| \hat{U}_{1,2} \,  {\cal O}(k_1^{-2}) \right| dk_1.
\end{equation*}
Since $\hat{U}_{1,2} = {\cal O}([\hat{u}^{(1)}]^3)$, then for shallow-water wave (long wavelength), it is sufficient to take any initial condition such that with $\hat{u}^{(1)} = {\cal O}(1)$ to guarantee that $\Omega_2$ is integrable or the integrand of $\Omega_2 \in L^2(\mathbb{R})$. This is also consistent with the imposed initial condition $\hat{u}_1 = \epsilon \hat{u}^{(1)}$.

Since we are interested in the region of the normalized wavenumber between the two wavenumber thresholds $k_s < k < k_d$, it is interesting to find possible values for these wavenumber thresholds $k_s$ and $k_d$. It is known in the literature that based on nonlinear, Stokes' wave theory, the characteristic for shallow-water waves or long waves is $kh \ll 1$, where $h$ is the mean water depth.
In our normalized notation, this will be $k \ll 1$. On the other hand, the characteristic for deep-water waves or short waves is $kh \gg 1$, or in normalized notation, $k \gg 1$.
Linear (Airy) wave theory provides a particular wavenumber range for surface gravity waves at intermediate depth, i.e. in normalized form given as $\pi/10 = k_s < k < k_d = \pi$~\citep{Dean91,Schwartz06,Mani12}.

However, an article by~\cite{Bird01} states that the wavenumber interval for intermediate-water wave (in normalized form) is $0.7 \lesssim k < 1.363$.
The upper wavenumber threshold of 1.363 suggests that the authors adopted it from the context of the stability of surface gravity waves, where the process of nonlinear focusing ceases to exist for sufficiently small water depth. This result was first discovered by~\cite{Benjamin67} and~\cite{Whitham74} when studying the instability of a uniform and finite-amplitude wave train; see also~\cite{Johnson77}. Since the approach implemented in this paper follows the Stokes expansion method, and not necessarily from the context of wave train stability, in the following, we will provide a possible (normalized) value for upper-bound wavenumber threshold $k_d$ as a transition from intermediate water-wave to deep-water wave.

In the literature, it is suggested but not proved that the wavenumber threshold $k_s$ for the transition from shallow-water wave to intermediate water-wave will be $\pi/10 < k_s \lesssim 0.7$~\citep{Dean91,Bird01}.
A potential value for wavenumber threshold for deep-water wave case $k_d$ can be found using an approximation for the integrand in the expression~\eqref{ndr1} or~\eqref{ndr2}.
We have $\Omega_0(2k_1) + 2 \Omega_0(k_1) < (2 + \sqrt{2}) \sqrt{k_1} < 4 \sqrt{k_1}$.
Using a hyperbolic counterpart of Wilker's inequality: $\tanh x/x > 2 - \left(\sinh x/x \right)^2, \; x \neq 0$~\citep{Wilker89,Zhu07} and
Huygen's inequality for the hyperbolic functions: $\tanh x/x > 3 - 2 \sinh x/x. \; x \neq 0$~\citep{Huygens,Neuman10}, it can be shown that the
integral~\eqref{ndr1} or~\eqref{ndr2} satisfies the inequality
\begin{equation*}
\int \frac{k_1 \hat{U}_{1,2}(k_1)\, dk_1}{|\Omega_0(2k_1) - 2 \Omega_0(k_1)|} < \int \frac{\hat{U}_{1,2}(k_1) {\cal{O}}(k_1^{3/2})}{2 k_1^2 - \sinh^2 k_1}\left(1 + \frac{1}{(3 k_1 - 2 \sinh k_1)^2} \right)\, dk_1 \qquad \textmd{for} \qquad k_s < k_1 < k_d.
\end{equation*}
This inequality will be satisfied for the imposed initial condition $\hat{u}^{(1)}$ approximately in the order of ${\cal O}(k^{-5/3})$ up to ${\cal O}(1)$.
Since the integrand's denominator of the right-hand side becomes singular when $2 k_1^2 - \sinh^2 k_1 = 0$,
this suggests a potential wavenumber threshold for deep-water wave $k_d \approx 1.49143$.
This finding shows that $k_d$ proposed in the context in this paper is consistent with the values of similar
wavenumber thresholds mentioned in the literature, i.e. $1.363 < k_d \approx 1.49143 < \pi$~\citep{Dean91,Bird01,Schwartz06,Mani12}.

\section{Conclusion and remark}

The KdV type of equation with exact dispersion property is revisited in this paper. The attempt to find a solution up to the third-order term has been implemented using the regular perturbation theory. Due to mode generation through nonlinearity, it turned out that the third-order term consists of resonant terms which grow linearly in time. The occurrence of these resonant terms can be prevented by implementing the method of strained parameters, also known as the Lindstedt--Poincar\'{e} technique. As a consequence, nonlinear dispersion relation terms for the second-order term are obtained which depend on the combinations of the wavenumbers mode generated through nonlinearity. The result from the validity analysis suggests that it is necessary to impose a particular order of the initial condition to guarantee the existence of the nonlinear dispersion relation. A simple analysis shows that a potential wavenumber threshold for the transition from the intermediate-water to the deep-water wave criteria lies consistently within the values found in the literature.

\section*{Acknowledgement}
{\small The author acknowledges Professor E. (Brenny) van Groesen (Universiteit Twente, The Netherlands and LabMath Indonesia), Professor Mark. J. Ablowitz (University of Colorado, Boulder), Professor Andy Chan Tak Yee (The University of Nottingham Malaysia Campus), Dr Gert Klopman (Universiteit Twente and Witteveen$+$Bos, The Netherlands), Dr Sivasankaran Sivanandam (National Taiwan University and University of Malaya, Malaysia) and Dr Mark Lawrence (Nazarbayev University, Kazakhstan) for many fruitful discussions as well as the anonymous referees for constructive suggestions for the improvement of this article. \par}

\section*{Funding}
{\small The research is supported by the project TWI.5374 of the Netherlands Organization of Scientific Research NWO, subdivision Applied Sciences STW, the New Researcher Fund NRF 5035-A2RL20 from the University of Nottingham, University Park Campus, UK and Malaysia Campus, Malaysia and the Seed Grant K$\Phi$-13/28 from the `Fund of Social Development' Corporate Fund and Central Research Office (CRO) of Nazarbayev University Research and Innovation System (NURIS), Kazakhstan}.

\end{document}